# Spatial Dynamics of Urban Growth Based on Entropy and Fractal Dimension

Yanguang Chen

(Department of Geography, College of Urban and Environmental Sciences, Peking University, Beijing 100871, P.R. China. E-mail: chenyg@pku.edu.cn)

**Abstract**: The fractal dimension growth of urban form can be described with sigmoid functions such as logistic function due to squashing effect. The sigmoid curves of fractal dimension suggest a type of spatial replacement dynamics of urban evolution. How to understand the underlying rationale of the fractal dimension curves is a pending problem. This study is based on two previous findings. First, normalized fractal dimension proved to equal normalized spatial entropy; second, a sigmoid function proceeds from an urban-rural interaction model. Defining urban space-filling measurement by spatial entropy, and defining rural space-filling measurement by information gain, we can construct a new urban-rural interaction and coupling model. From this model, we can derive the logistic equation of fractal dimension growth strictly. This indicates that urban growth results from the unity of opposites between spatial entropy increase and information increase. In a city, an increase in spatial entropy is accompanied by a decrease in urban land availability. An inference is that urban growth struggles between the force of urban space-filling and the force of urban space-saving. This work presents a set of urban models of spatial dynamics, which help us understand urban evolution from the angle of view of entropy and fractals.

## 1. Introduction

In order to understand how cities evolve, we must describe urban form and growth using mathematics and measurement. Effective description is the precondition of deep understanding of city development. Traditional mathematical approaches are mainly based on the concept of characteristic scales, while many aspects of cities are scale free and cannot be effectively described by the conventional mathematic methods. Fractal geometry provides a powerful tool to make scaling analysis of cities and systems of cities (Batty, 2008; Batty and Longley, 1994; Frankhauser, 1994; Frankhauser, 1998). Cities as fractals bear no characteristic scales, but fractal dimension represents the characteristic values of scaling transform and can be used to characterize urban patterns and processes. Fractal parameters have clear lower limit and upper limit. Suppose a fractal city is defined in a 2-dimensional space, the fractal dimension $D$ values ranges from 0 to 2 in theory ($0<D<2$). The lower and upper limits of fractal parameters result in a squashing effect, and a fractal dimension curve of urban growth is supposed to be a sigmoid curve coming between the two limit lines (Chen, 2012).The most typical sigmoid function in the squashing function family is the well-known logistic function, which can be employed to model fractal dimension growth of the cities in developed countries and regions. The logistic growth of fractal parameters of urban form indicates spatial replacement dynamics of urban and rural interaction. However, how to reveal the underlying rationale of this replacement dynamics is still a pending question.

Fractal dimension is a sort of space filling index, indicating the degree of spatial occupancy. A fractal dimension value can reflect the spatial feature of urban form, and a fractal dimension curve based on a set of fractal dimension values in a time series can mirror urban growth. On the other, the models of fractal dimension curve can be used to predict urban growth and explore the mechanism of urban evolution. A sigmoid function suggests complex dynamics (May, 1976), and is often associated with a kind of interaction models (Chen, 2009; Karmeshu, 1988).If we can derive the sigmoid function of fractal dimension curves from some spatial models, we will approach the general principle of urban evolution. The key lies in finding the basic spatial measurements supporting fractal parameters. Based on the basic measurements, we can construct the model of spatial dynamics. In fact, two basic measurements have been found: one is spatial entropy, and the other, information gain (Batty, 1976).The spatial entropy used to be calculated by systems of geographical zones. Substituting functional boxes for geographical zones, we can relate spatial entropy with fractal dimension. Fractal dimension is actually defined on the basis of entropy function. Hausdorff dimension proved to be equivalent to Shannon's information entropy (Ryabko, 1986). By means of association of entropy with fractal dimension, a new way of modeling spatial dynamics can be found to explore the rationale of urban evolution.

This paper presents a new model of spatial dynamics based on spatial entropy and fractal dimension. From this model, the logistic function of fractal dimension curve can be strictly derived. Thus the intrinsic relations between fractal dimension, fractal dimension curve, and urban-rural interaction of cities can be brought to light. The study rests with a normative framework, two basic concepts, and three discoveries. The normative framework is functional box-counting method, and the two concepts are spatial entropy and information gain. The three previous findings are as follows. First, normalized fractal dimension is equal to normalized spatial entropy; second, squashing effect leads to a sigmoid curve of fractal dimension; third, the sigmoid function proceeds from an urban-rural interaction model. The rest parts are arranged as below: In Section 2, based on the previous findings, the results of mathematical modeling are presented; in Section 3, based on the study of Murcio *et al* (2015), empirical evidence from London is shown to support the theoretical results; in Section 4, the related questions are discussed. Finally, the main conclusions are drawn from the theoretical results and empirical analysis.

## 2. Theoretical results

### 2.1 Logistic equation of fractal dimension curve

The curves of fractal dimension growth are termed fractal dimension curves, which can be described with sigmoid functions. The fractal dimension growth of the cities in developed countries can be modeled by the common Boltzmann equation, while the fractal dimension curves of cities in developing countries can be modeled by quadratic Boltzmann equation. The common Boltzmann equation of fractal dimension curves is as follows (Chen, 2012)

$$D(t) = D_{\min} + \frac{D_{\max} - D_{\min}}{1+[\frac{D_{\max} - D_{(0)}}{D_{(0)} - D_{\min}}]e^{-kt}} = D_{\min} + \frac{D_{\max} - D_{\min}}{1+\exp(-\frac{t-t_0}{p})}, \quad (1)$$

where $D(t)$ refers to the fractal dimension of urban form in time $t$, $D_{(0)}$ to the fractal dimension in the initial time, $D_{\max}$ is the capacity parameter of fractal dimension (the upper limit), $D_{\min}$ is the minimum fractal dimension value (the lower limit), $k$ represents the initial growth rate, $p=1/k$ is a scaling parameter associated with $k$, and $t_0=\ln[(D_{\max}-D_{(0)})/(D_{(0)}-D_{\min})]^p$ is a temporal translational parameter indicative of a critical time. Empirically, $D_{max} \leq 2$, $D_{min} \geq 0$. For the normalized fractal dimension, equation (1) can be rewritten as a logistic function such as

$$D^*(t) = \frac{D(t) - D_{min}}{D_{max} - D_{min}} = \frac{1}{1 + (1/D^*_{(0)} - 1)e^{-kt}}, \qquad (2)$$

where $D^*_{(0)} = (D_{(0)} - D_{min})/(D_{max} - D_{min})$ denotes the normalized value of the original fractal dimension. Derivative of equation (2) with respect to time $t$ is a logistic equation

$$\frac{dD^*(t)}{dt} = kD^*(t)[1 - D^*(t)], \qquad (3)$$

where d refers to derivative. In other words, equation (2) is a special solution to equation (3). Discretizing equation (3) yields a 1-dimensional logistic map. A 1-dimensional map is always associated with a 2-dimensional map (Chen, 2009). This suggests that we can find a two-element interaction model to explain the logistic growth of fractal dimension.

Urban growth is a dynamic process of space filling and spatial replacement. Rural space is replaced by urban space, and natural landscape is replaced by human landscape (Chen, 2012; Chen, 2014). This process is associated with urbanization. City development can be illustrated with a regular growing fractal, which was proposed by Jullien and Botet (1987) and popularized by Vicsek (1989) (Figure 1). It bears an analogy with urban growth, and can be used to model urban form (Batty and Longley, 1994; Frankhauser, 1998; Longley *et al*, 1991; White and Engelen, 1993). In fact, introducing chance factors to the regular fractal yields random fractals such as diffusion-limited aggregation (DLA) model (Witten and Sanders, 1981; Witten and Sanders, 1983). DLA model has been employed to simulate urban growth (Batty *et al*, 1989; Fotheringham *et al*, 1989). Suppose that the normalized fractal dimension equals the ratio of urban space to total space, that is

$$D^*(t) = \frac{U(t)}{U(t) + V(t)}. \qquad (4)$$

If we validate equation (4), then we will be able to interpret the logistic growth of fractal dimension of urban form using a model of spatial replacement dynamics (Chen, 2012). The model can be expressed with a pair of differential equations as follows

$$\frac{dU(t)}{dt} = \alpha U(t) + \beta \frac{U(t)V(t)}{U(t) + V(t)}, \qquad (5)$$

$$\frac{dV(t)}{dt} = \lambda V(t) - \beta \frac{U(t)V(t)}{U(t)+V(t)}, \tag{6}$$

where $U(t)$ refers to a urban space measurement of time $t$, and $V(t)$ to the corresponding rural space measurement, $\alpha$, $\beta$, and $\lambda$ are parameters. Based on equation (4), equation (3) can be derived from combination of equations (5) and (6). Thus, we can understand equation (3) through the interaction between $U(t)$ and $V(t)$. Now, a problem arises: how to define the two spatial measurements: $U(t)$ and $V(t)$? This is one of important tasks of this study.

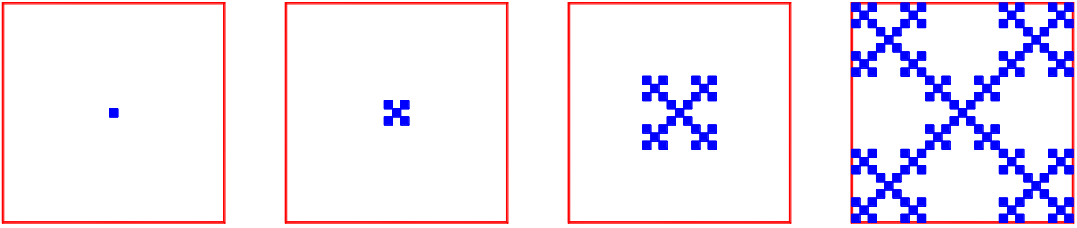

**Figure 1A sketch map of fractal growth, space filling, and spatial replacement (the first four steps)**

[**Note**: If a city grows up from a small human settlement such as a village, it can be modeled with this kind of growing fractals. A simple model of fractal growth is actually a metaphor of city development.]

## 2.2 Spatial entropy and information gain

Fractal dimension is a space filling measure indicative of complex patterns, and fractal dimension growth of urban form suggests urban space filling processes. A finding in this work is that the spatial measurements $U(t)$ and $V(t)$ can be defined using spatial entropy. This will be illuminated step by step. In fact, fractal dimension formulae are based on entropy functions. Entropy proved to be the measure of multiscale complexity (Bar-Yam, 2004a; Bar-Yam, 2004b). Spatial entropy can be associated with multifractal parameters by box-counting method. The generalized correlation dimension of multifractals is defined by Renyi's entropy (Renyi, 1961), which is as below

$$M_q = -\frac{1}{q-1} \ln \sum_{i=1}^{N} P_i^q, \tag{7}$$

where $q$ denotes the order of moment, $M_q$ refers to Renyi's entropy of order $q$, $P_i$ is the growth probability of the $i$th fractal copy, $N$ is the number of fractal copies ($i=1,2,3,\ldots,N$). A fractal copy can be regarded as a fractal unit in the fractal set. In the spatial measurement of random fractals by means of box-counting method, the fractal copies are substituted with boxes. The information gain can be defined as (Batty, 1974)

$$G_q = M_{\max} - M_q = \ln N_T + \frac{1}{q-1} \ln \sum_{i=1}^{N} P_i^q, \qquad (8)$$

where $G_q$ denotes the information gain of order $q$, $M_{\max}=\ln N_T$ refers to the maximum entropy, $N_T$ is total number of possible fractal units, including the fractal copies in the fractal set and the spatial units in the fractal complement (Figure 2). If $q=1$ and $N_T=N$, then the information gain becomes the $H$ quantity in the theory of dissipative structure (Prigogine and Stengers, 1984).

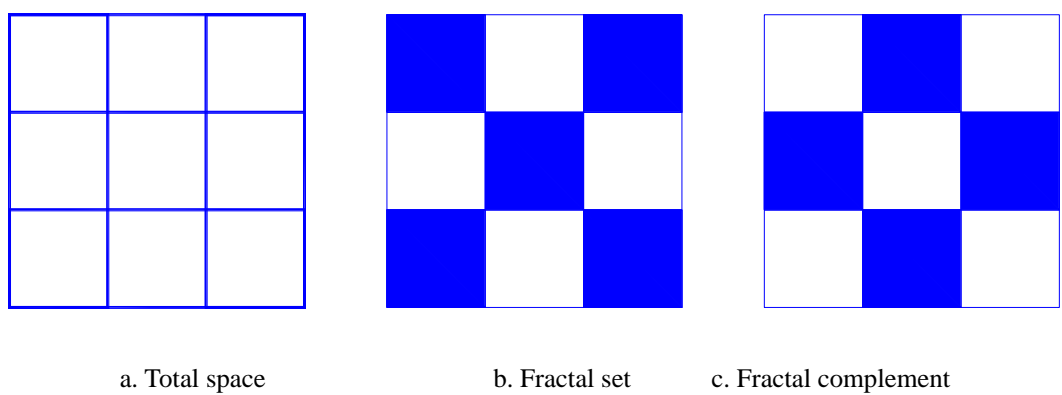

a. Total space    b. Fractal set    c. Fractal complement

**Figure 2 A set of schematic diagrams about total space, fractal set, and fractal complement (the second step)**

[**Note**: (a) The total number of possible fractal units is $N_T=9$, the maximum entropy is $M_{\max}=\ln(9)=2.1972$ nat; (b) The actual number of fractal units is $N=5$, the actual entropy is $M_q=\ln(5)=1.6094$ nat; (c) The number of residual spatial units is $N_T-N=4$, and the information gain is $G_q= M_{\max}-M_q=0.5878$ nat rather than $\ln(4)=1.3863$ nat.]

For fractal cities, spatial entropy can be associated with fractal dimension with functional box-counting method. A fractal city is mainly defined in a 2-dimensional space (Batty and Longley, 1994). In other words, the embedding space of city fractals is Euclidean space with a dimension $d=2$. There are two orthogonal directions in a 2-dimensional embedding space. Thus the total possible urban space can be theoretically defined in a square or rectangular area, which is termed *measure area* in mathematics. On the other hand, the Lebesgue measure of a fractal proved to be zero. So we have $D_{\max}=2$ and $D_{\min}=0$ in theory. As a result, the normalized fractal dimension can be expressed as

$$D_q^* = \frac{D_q - D_{\min}}{D_{\max} - D_{\min}} = \frac{D_q}{D_{\max}}, \tag{9}$$

where $D_q$ is the generalized correlation dimension of order $q$. Based on box-counting method, the normalized spatial entropy is equal to the normalized fractal dimension (Chen, 2016), that is

$$J_q = \frac{M_q}{M_{\max}} = -\frac{1}{q-1} \ln \sum_{i=1}^{N} P_i^q / \ln N_T = \frac{D_q}{D_{\max}} = D_q^*. \tag{10}$$

where $J_q$ denotes the entropy ratio. Equation (10) indicates

$$D_q = D_{\max} D_q^* = d D_q^* = d J_q, \tag{11}$$

where $d = D_{\max} = 2$. In equation (4), let $U_q = M_q$, $V_q = G_q = M_{\max} - M_q$, then $M_{\max} = U_q + V_q = M_q + G_q$. Thus we have

$$D_q^*(t) = \frac{U_q(t)}{U_q(t) + V_q(t)} = \frac{M_q(t)}{M_{\max}(t)} = 1 - \frac{G_q(t)}{M_{\max}(t)}. \tag{12}$$

By means of equation (12), we can derive equation (3) from equations (5) and (6); and by equation (11), we can return the normalized fractal dimension ($D_q^*$) to the common fractal dimension ($D_q$). In fact, using equation (11), we can generalize multifractal measures to describe both scale-free distributions and the distributions with characteristic scales.

## 2.3 Entropy evolution leads to fractal dimension curves

The spatial replacement dynamics of urban growth is in essence a process of interaction between spatial entropy and information gain. In multifractal theory, the information gain can be regarded as general $H$ quantity. If the spatial entropy $M_q$ is employed to measure the urban space filling extent $U_q$, then the information gain $G_q$ can be used to measure the urban space saving extent or rural anti-filling extent $V_q$. Thus equations (5) and (6) can be rewritten as follows

$$\frac{dM_q(t)}{dt} = \alpha M_q(t) + \beta \frac{M_q(t) G_q(t)}{M_{\max}(t)}, \tag{13}$$

$$\frac{dG_q(t)}{dt} = \lambda G_q(t) - \beta \frac{M_q(t) G_q(t)}{M_{\max}(t)}, \tag{14}$$

which reflect the coupling relationships and interaction between urban space and rural space. Taking derivative of equation (12) yields

$$\frac{dD_q^*(t)}{dt} = \frac{dM_q(t)/dt}{M_{\max}(t)} - \frac{M_q(t)}{M_{\max}^2(t)}\left[\frac{dM_q(t)}{dt} + \frac{dG_q(t)}{dt}\right]. \tag{15}$$

Substituting equations (13) and (14) into equation (15) yields

$$\frac{dD_q^*(t)}{dt} = \beta \frac{M_q(t)G_q(t)}{M_{\max}^2(t)} + \frac{M_q(t)}{M_{\max}(t)}\left[\alpha - \alpha\frac{M_q(t)}{M_{\max}(t)} - \lambda\frac{G_q(t)}{M_{\max}(t)}\right]. \tag{16}$$

Further substituting equation (12) into equation (16) yields

$$\frac{dD_q^*(t)}{dt} = \beta \frac{M_q(t)}{M_{\max}(t)}[1 - \frac{M_q(t)}{M_{\max}(t)}] + \frac{M_q(t)}{M_{\max}(t)}[\alpha(1 - \frac{M_q(t)}{M_{\max}(t)}) - \lambda(1 - \frac{M_q(t)}{M_{\max}(t)})], \tag{17}$$

which can be reduced to

$$\frac{dD_q^*(t)}{dt} = (\alpha + \beta - \lambda)D_q^*(t)[1 - D_q^*(t)]. \tag{18}$$

Comparing equations (18) with equation (3) shows a parameter relation such as $k=\alpha+\beta-\lambda$. Equation (18) multiplied by $D_{\max}$ yields

$$\frac{dD_q(t)}{dt} = kD_q(t)[1 - \frac{D_q(t)}{D_{\max}}], \tag{19}$$

which is the logistic equation of general fractal dimension growth of urban form. Apparently, substituting equation (11) into equation (19) yields equation (18). According to the process of mathematical derivation, equations (13) and (14) are equivalent to equation (18), which is in turn equivalent to equation (19). The solution to equation (19) is the general logistic function of fractal dimension curve, that is

$$D_q(t) = \frac{D_{\max}}{1 + (D_{\max}/D_{q(0)} - 1)e^{-kt}}, \tag{20}$$

which can be testified by observational data of fractal dimension of urban form in the real world.

## 3. Materials and method

### 3.1 Empirical analysis

In urban studies, a complete research comprises three analytical processes. The first is theoretical modeling, the second is empirical analysis, and the third is computer simulation or numerical experiment. Where empirical analysis is concerned, many direct and circumstantial evidences can be found to support the inference that the fractal dimension curves of urban growth of European and American cities conform to the common logistic function. In fact, a number of cities in developed countries or regions have been investigated and the fractal dimension datasets are available (Chen, 2012; Chen, 2014; Chen and Huang, 2016). Batty and Longley (1994) have published a dataset of fractal dimension time series of London(partial data came from Frankhauser, 1994), Benguigui *et al* (2000) have published several datasets of fractal dimension time series of the morphology of the Tel-Aviv metropolis, and Shen (2002) has presented a fractal dimension time series of Baltimore's urban form. The common logistic function can be fitted to the above-mentioned fractal dimension series. Sun and Southworth (2013) have displayed a number of fractal dimension time series of the cities and towns in the developed areas of Amazon tri‑national frontier regions. All these fractal dimension series but one can be fitted with the logistic function (Chen, 2014). Recently, Murcio *et al* (2015) calculated the generalized fractal dimension time series of London (Table 1). On the whole, the London's growth curves of capacity dimension, information dimension, and correlation dimension accord with the common logistic curve, which will be emphatically investigated in this section.

Table 1 The typical cities and towns with fractal dimension curve of logistic growth

| City | Country | Period | Data point | Source |
| --- | --- | --- | --- | --- |
| London | UK | 1820-1981 | 8 years | Batty and Longley, 1994; Frankhauser, 1994 |
| Tel-Aviv | Israel | 1935-1991 | 7-8 years | Benguigui *et al*, 2000 |
| Baltimore | USA | 1792-1992 | 12 years | Shen, 2002 |
| Amazon tri-national cities and towns | Brazil, Peru, Bolivia | 1986-2010 | 6 years | Sun and Southworth, 2013 |
| London | UK | 1786-2010 | 9 years | Murcio *et al*, 2015 |

The fractal dimension datasets of London can be employed to make a simple case study. Compared with the previous studies on monofractals (unifractals), the work of Murcio *et al* (2015) is on multifractal cities (Table 2).The fractal dimension time series can be described with the logistic model, equation (20). The parameters can be estimated using the least squares calculation. Based on the confidence level of 95%, the results are satisfying and acceptable (Table 3). In contrast, fitting the quadratic logistic function to the datasets shown in Table 2 cannot yield satisfying results. The logistic models of three types of fractal dimension curves can be given by the estimated parameter values. For capacity dimension $D_0(q=0)$, the logistic growth model is as below

$$\hat{D}_0(t) = \frac{2}{1+0.1251e^{-0.0036t}}.$$

The goodness of fit is about $R^2=0.9395$. The hat "^" implies predicted value. Similarly hereinafter.

For information dimension $D_1$ ($q=1$), the logistic model is in the following form

$$\hat{D}_1(t) = \frac{2}{1+0.1458e^{-0.0040t}}.$$

The goodness of fit is around $R^2=0.9314$. For correlation dimension $D_2$ ($q=2$), the logistic model is as follows

$$\hat{D}_2(t) = \frac{2}{1+0.1667e^{-0.0046t}}.$$

The goodness of fit is about $R^2=0.9197$. For all these models, the maximum dimension values are $D_{max}=2$. By means of these models, we can predict the three types of fractal dimension. The observational data and the calculated values are generally consistent with one another (Figure 3).

**Table 2 The capacity dimension, information dimension, and correlation dimension of London's urban form (1786-2010)**

| Year $n$ | Moment order $q$ | Time $t$ | Fractal dimension | | |
|---|---|---|---|---|---|
| | | | $D_0$ | $D_1$ | $D_2$ |
| 1786 | [−5.00,14.25] | 0 | 1.7959 | 1.7697 | 1.7467 |
| 1830 | [−7.00,13.25] | 44 | 1.7926 | 1.7698 | 1.7448 |
| 1880 | [−5.75,14.50] | 94 | 1.8196 | 1.7950 | 1.7722 |
| 1900 | [−4.75,19.00] | 114 | 1.8434 | 1.8252 | 1.8109 |
| 1920 | [−4.25,17.75] | 134 | 1.8602 | 1.8431 | 1.8322 |
| 1940 | [−5.25,18.75] | 154 | 1.8699 | 1.8619 | 1.8586 |
| 1965 | [−5.25,16.00] | 179 | 1.8850 | 1.8803 | 1.8780 |
| 1990 | [−4.50,15.00] | 204 | 1.8851 | 1.8793 | 1.8766 |
| 2010 | [−4.50,16.00] | 224 | 1.8913 | 1.8858 | 1.8842 |

**Source**: Murcio et al, 2015.

**Table 3 The logistic model parameters and the goodness of fit on London's urban growth (1786-2010)**

| Parameter /Statistic | Fractal dimension models | | |
|---|---|---|---|
| | Capacity dimension $D_0$ | Information dimension $D_1$ | Correlation dimension $D_2$ |
| $D_{max}$ | 2 | 2 | 2 |
| $k$ | 0.0036 | 0.0040 | 0.0046 |
| $a=D_{max}/D_q(0)-1$ | 0.1251 | 0.1458 | 0.1667 |
| $R^2$ | 0.9395 | 0.9314 | 0.9197 |

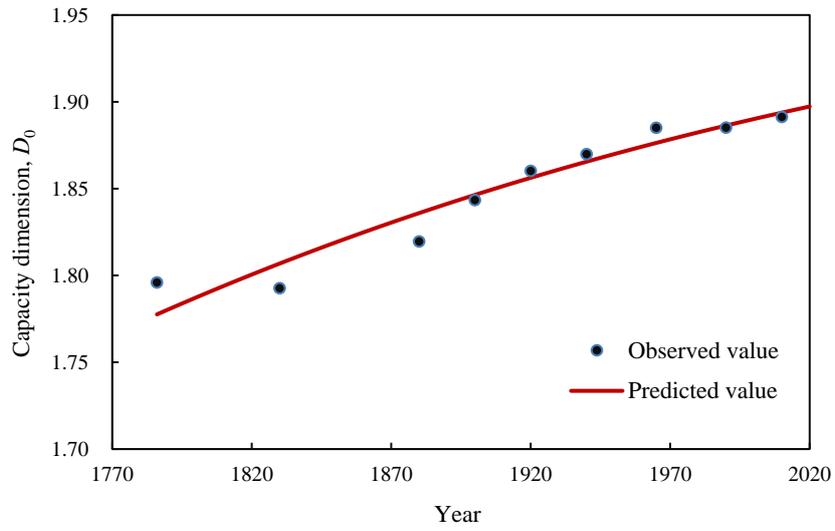

a. Capacity dimension ($1.75<D_0<2.0$)

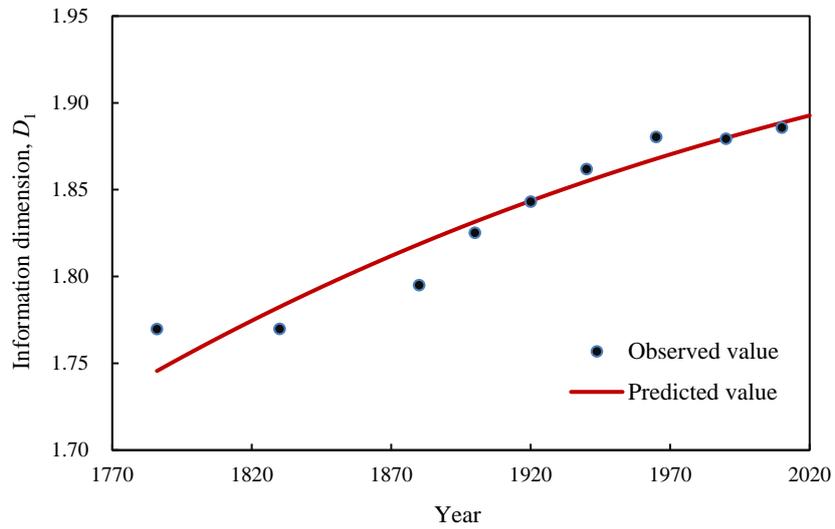

b. Information dimension ($1.7<D_1<2$)

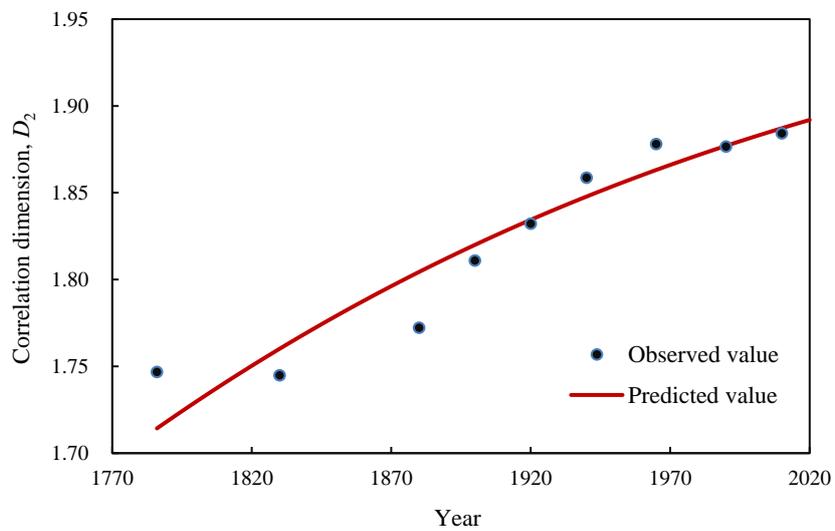

c. Correlation dimension ($1.7<D_2<2$)

**Figure 3 Three logistic curves of fractal dimension growth of London's urban form (1786-2020)**

[**Note**: In the early years, i.e., from 1786 to 1880, the caliber of the data seems to be not consistent with that in recent years. As a result, the scatter points do not well match the trend lines.]

### 3.2 Numerical experiment

Numerical iteration process can be employed to reveal the mathematical relation between the logistic function and the urban-rural interaction model. Numerical method is an effective approach to finding the solutions of nonlinear equations (Bak, 1996). As shown above, equations (2) and (20) can be derived from equations (13) and (14). If the sigmoid curves given by equations (2) and (20) are consistent with the curves generating by equations (13) and (14), the derivation of logistic growth from urban-rural replacement dynamics will be supported by the numerical analysis. Discretizing equations (13) and (14) yields a 2-dimensional map as below:

$$M_{t+1} = (1+\alpha)M_t + \beta \frac{M_t G_t}{M_t + G_t}, \tag{21}$$

$$G_{t+1} = (1+\lambda)G_t - \beta \frac{M_t G_t}{M_t + G_t}, \tag{22}$$

where $M_t$ and $G_t$ represent the discrete forms of $M(t)$ and $G(t)$, respectively, and $M_t+G_t=M_{\max, t}$, in which $M_{\max, t}$ is the discrete expression of $M_{\max}(t)$. If we can utilize equations (21) and (22) to generate logistic curves, which can be fitted to the observational data of fractal dimension shown in Table 2, the relationships between the logistic function and the urban-rural replacement model will be supported by numerical experiments.

The primary step of numerical iteration analysis is to set proper initial values of fractal dimension and the parameter values of urban-rural replacement model. Suppose that the city originated from a hamlet in a large region, and the initial fractal dimension is close to zero. The parameter values of equations (21) and (22) can be selected by testing again and again in terms of the relation $k=\alpha+\beta-\lambda$, in which the $k$ values are displayed in Table 3. The initial values of fractal dimension and the parameter values of the discrete dynamics model are shown in Table 4. The main findings from this numerical analysis are as follows. **First, the 2-dimensional map from the urban-rural replacement model does generate sigmoid curves.** These curves are generally consistent with the observational data of multifractal parameters (Figure 4). **Second, the parameter values must be subject to the constraint conditions.** The constraints are $\alpha+\beta-\lambda=k$ and $k_0 \geq k_1 \geq k_2$, where $k_0$, $k_1$, and $k_2$ represents the $k$ values for capacity dimension $D_0$, information dimension $D_1$, and correlation dimension $D_2$. As we know, the generalized correlation function is a monotone decreasing function, and $D_0 \geq D_1 \geq D_2$. If $k_0 < k_1 < k_2$, then $D_0 < D_1 < D_2$, which indicates an illogical relation. This suggests that the $k$ values shown in Table 3 are problematic numbers. **Third, there are some flaws in the relationships between different parameter values of the discrete dynamic model.** The 2-dimensional map can generate satisfying sigmoid curve for capacity dimension, but it is difficult to yield satisfying correlation dimension curve. In other word, the calculated values of correlation dimension are not very consistent with the corresponding observational data. The effect of information dimension curves comes between capacity dimension and correlation dimension. The goodness of numerical simulation corresponds to the goodness of logistic fit, $R^2$ (see Table 3).

Table 4 The initial values of spatial entropy, information gain, fractal dimension, and parameter values of the urban-rural replacement model

| Fractal dimension | Initial values for entropy and dimension | | | | Parameter values for dynamics model | | | |
|---|---|---|---|---|---|---|---|---|
| | $M_0$ | $G_0$ | $D^*_{(0)}$ | $D_{(0)}$ | $\alpha$ | $\beta$ | $\lambda$ | $k$ |
| $D_0$ | 0.1000 | 0.9000 | 0.1000 | 0.2000 | 0.0028 | 0.0023 | 0.0015 | 0.0036 |
| $D_1$ | 0.0900 | 0.8800 | 0.0928 | 0.1856 | 0.0029 | 0.0023 | 0.0016 | 0.0036 |
| $D_2$ | 0.0850 | 0.8700 | 0.0890 | 0.1780 | 0.0030 | 0.0023 | 0.0017 | 0.0036 |

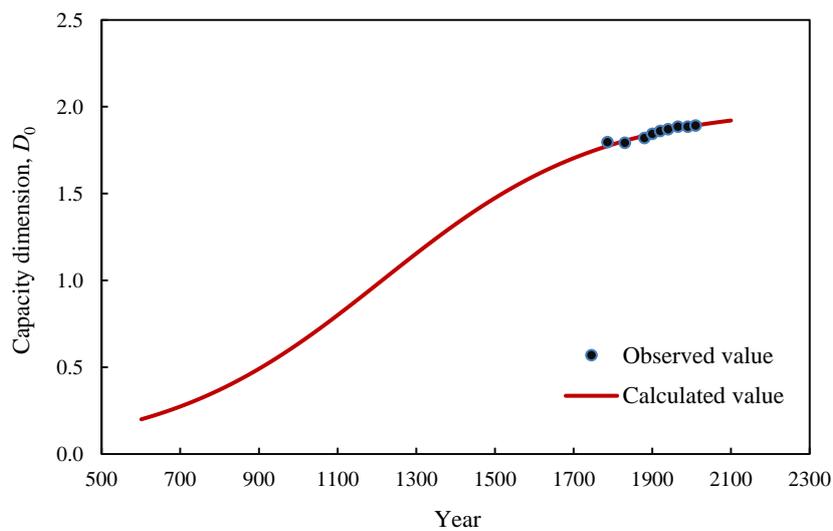

a. Capacity dimension ($0.0<D_0<2.0$)

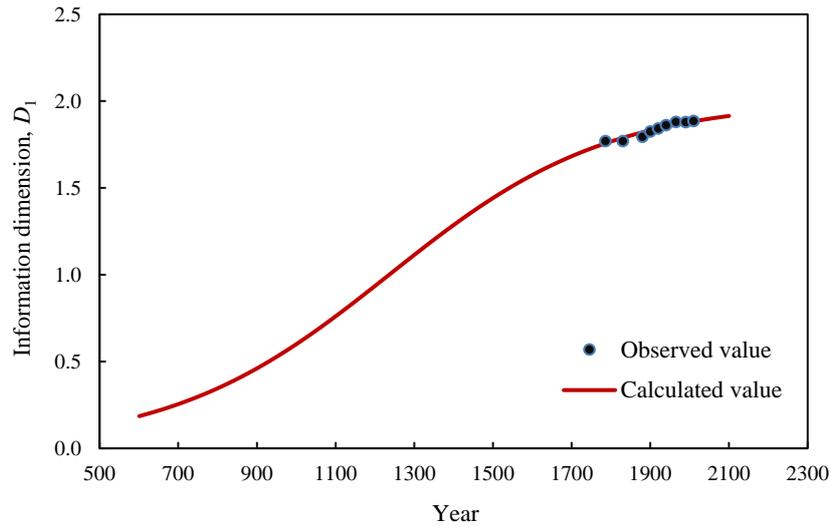

b. Information dimension ($0.0<D_1<2$)

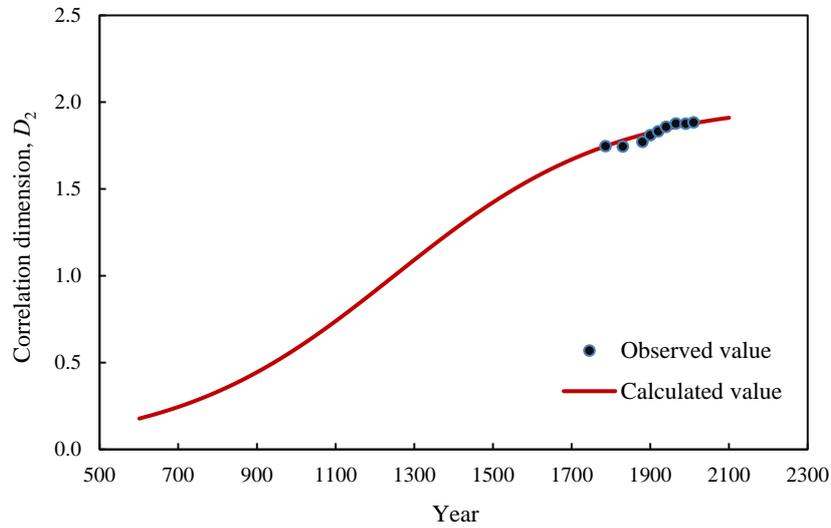

c. Correlation dimension ($1.7<D_2<2$)

**Figure 4 Modeling three logistic curves of fractal dimension growth of London's urban form with**

**2-dimensional urban-rural replacement map (600-2100)**

[**Note**: The time scale is extended so that the trend of logistic growth can be clearly displayed.]

## 4. Discussion

There is an analogy between the thermodynamic entropy in physics and the spatial (information) entropy in geography. In physics, entropy is a measure of the unavailability of a system's energy to do work. For a closed system, an increase in entropy is accompanied by a decrease in energy availability. In a wider sense, entropy can be interpreted as a measure of disorder: the higher the entropy, the greater the disorder. In urban studies, spatial entropy is a measure of the unavailability of a city's land. In a confined city, an increase in spatial entropy is accompanied by a decrease in urban land availability. For a given urban field, the maximum spatial entropy is determinate, and the information gain can be computed. If spatial entropy represents the force of urban growth and space filling, the information gain represents the opposing force indicating urban antigrowth and anti-filling of space (Table 4). The competition between spatial entropy and information gain leads to spatial complexity. According to equations (13) and (14), the growth rate of spatial entropy, $dM_q(t)/dt$, is proportional to the current value of spatial entropy, $M_q(t)$, and the coupling between entropy and information gain, but not directly related to information gain; the growth rate of information gain, $dG_q(t)/dt$, is proportional to the current value of information gain, $G_q(t)$, and the coupling between entropy and information gain, but not directly related to the spatial entropy. Form equations (13) and (14) it follows the logistic equation of fractal dimension growth of urban form. This suggests that the interaction between spatial entropy and information gain results in the logistic growth of fractal dimension. The dynamics of urban-rural space replacement rests with the coupling relation and unity of opposites between spatial entropy and information gain.

Table 4 Comparison between two urban spatial measurements with unity of opposites

| Item | Urban space-filling extent $U_q$ | Urban space-saving extent $V_q$ |
| --- | --- | --- |
| Measurement | Spatial entropy $M_q$ | Information gain $G_q$ |
| Urban evolution | Entropy increase | Information increase |
| Geographical meaning | Decrease in land availability | Decrease of land waste |
| Physical meaning | Increase of spatial disorder | Increase of spatial order |

A city seems to be an enormous contradiction coming between spatial order and chaos. On the one hand, the logistic function of fractal dimension curves suggests spatial entropy increase. The increase of spatial entropy results in spatial homogeneity. In fact, fractal cities indicate spatial scaling (Batty and Longley, 1994), and single scaling process implies homogeneity (Feder, 1988). On the other hand, urban form proved to be multifractal patterns (Ariza-Villaverde *et al*, 2013; Chen and Wang, 2013; Murcio *et al*, 2015). Multifractal scaling provides an effective quantitative description of a variety of heterogeneous phenomena (Stanley and Meakin, 1988).Multifractal urban patterns represent typical spatial heterogeneity because the fractal dimension values of different parts are different from one another. The unity of opposites between spatial entropy increase and spatial heterogeneity of urban growth and form can be interpreted now. In light of above-shown mathematical derivation, two forces of urban evolution have been brought to light: one is urban growth force indicative of sprawl and space filling, and the other is the antigrowth force indicative of constraint and space saving. The former can be measured with spatial entropy, and the latter can be measured with information gain (Figure 5). The spatial entropy increase is constrained by the power of information gain of cities. The paradoxical movement of two different forces brings about spatial complexity and heterogeneity.

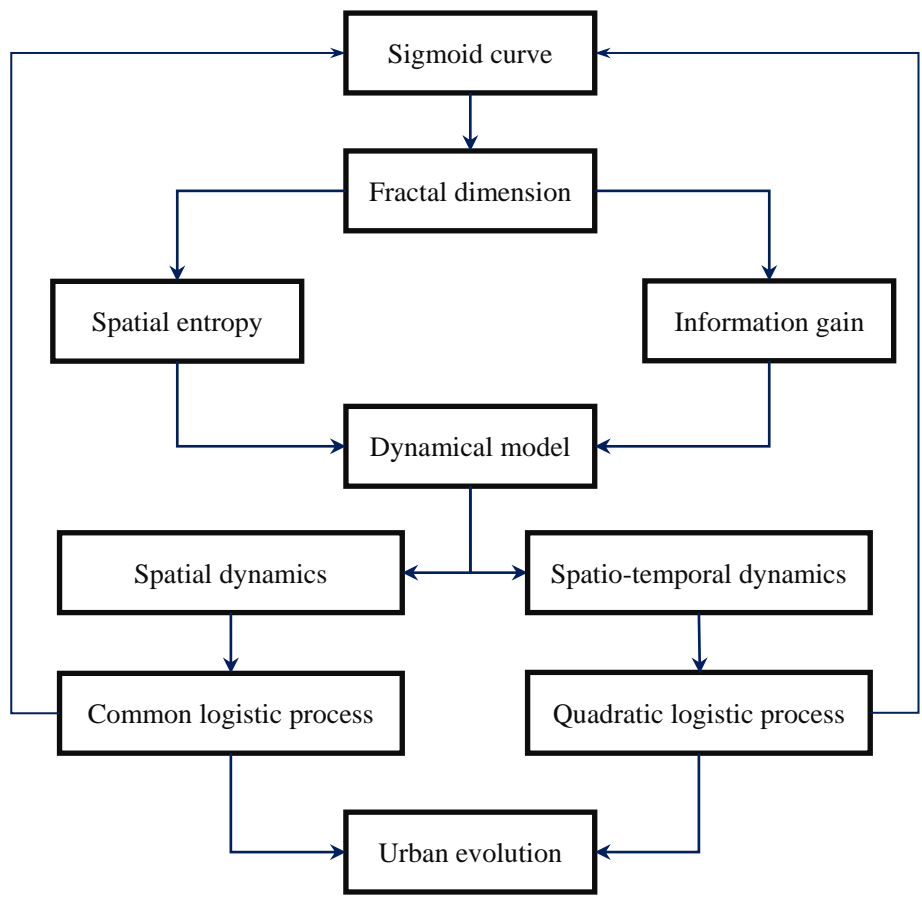

**Figure 5 A schematic diagram of modeling spatial dynamics of urban evolution using fractal dimension and spatial entropy**

The main academic contribution of this work is the demonstration that the logistic fractal dimension curve of urban growth can be derived from the model of interaction between spatial entropy and information gain. In previous literature, two findings have been made. One is that the fractal dimension of urban form takes on squashing effect and logistic growth, and the other is that the logistic fractal dimension curve can be interpreted by the urban-rural replacement dynamics (Chen, 2012; Chen, 2014). However, how to define the urban space-filing extent and the corresponding space-saving extent (rural space-filling extent) is a pending problem. This study is based on a significant finding that the normalized fractal dimension equals the normalized spatial entropy. If the urban space-filling extent is measured with spatial entropy, then the urban space-saving extent can be measured by the corresponding information gain. Thus the model of urban-rural replacement dynamics can be substituted by the model of interaction between entropy and information gain. As a result, the logistic function of fractal dimension growth of urban form can be derived out of the spatial entropy interaction model naturally and gracefully.

The logistic model of fractal dimension curves can be generalized to more universal form. The common logistic function can be employed to model fractal dimension growth of the urban form in developed countries and regions. However, the limitation of this model is clear. It is not suitable for the fractal dimension curves of the cities in developing countries such as China. The fractal dimension growth of urban form in China can be described with quadratic Boltzmann's equation such as

$$D(t) = D_{\min} + \frac{D_{\max} - D_{\min}}{1+[\frac{D_{\max} - D_{(0)}}{D_{(0)} - D_{\min}}]e^{-(kt)^2}} = D_{\min} + \frac{D_{\max} - D_{\min}}{1+\exp(-\frac{t^2 - t_0^2}{p^2})}. \quad (23)$$

Based on normalized fractal dimension, equation (23) changes to a quadratic logistic function as below

$$D^*(t) = \frac{D(t) - D_{\min}}{D_{\max} - D_{\min}} = \frac{1}{1+(1/D_{(0)}^* - 1)e^{-(kt)^2}}. \quad (24)$$

Accordingly, the model of interaction between entropy and information gain should be replaced by the following dynamics model

$$\frac{dM_q(t)}{dt} = t[\alpha M_q(t) + \beta \frac{M_q(t)G_q(t)}{M_{\max}(t)}], \quad (25)$$

$$\frac{\mathrm{d}G_q(t)}{\mathrm{d}t} = t[\lambda G_q(t) - \beta \frac{M_q(t)G_q(t)}{M_{\max}(t)}]. \tag{26}$$

From equations (24) and (25) it follows a quadratic logistic equation as follows

$$\frac{\mathrm{d}D^*(t)}{\mathrm{d}t} = 2k^2 t D^*(t)[1 - D^*(t)]. \tag{27}$$

The notation in equations (23) to (27) is the same as that in equations (1) to (6). A special solution to equation (27) is just equation (24). This suggests a difference of spatial dynamics of urban evolution between developed countries and developing countries. Time factor (*t*) plays an important role in the process of city development of developing countries. In contrast, in developed counties, the fractal dimension growth rate of urban form does not depend on time itself. An empirical analysis has been made by Chen and Huang (2016) on the quadratic logistic function of fractal dimension curves. Based on two types of fractal dimension curves, a general Boltzmann equation can be proposed as below:

$$D(t) = D_{\min} + \frac{D_{\max} - D_{\min}}{1 + [\frac{D_{\max} - D_{(0)}}{D_{(0)} - D_{\min}}]e^{-(kt)^\sigma}} = D_{\min} + \frac{D_{\max} - D_{\min}}{1 + \exp(-\frac{t^\sigma - t_0^\sigma}{p^\sigma})}, \tag{28}$$

in which the latent scaling parameter $\sigma$ comes between 1/2 and 2. If $D_{\min}$=0 as given, equation (28) will change to a generalized logistic function

$$D(t) = \frac{D_{\max}}{1 + (D_{\max}/D_{(0)} - 1)e^{-(kt)^\sigma}}, \tag{29}$$

which is a fractional logistic function of fractal dimension growth. When $\sigma$=1, we have a common logistic function of fractal dimension curve; when $\sigma$=2, we have a quadratic logistic function. Sometimes, the parameter $\sigma$ takes on a fraction such as 1.5. This suggests that the cities in developed countries and those in developing countries following the same mathematical law, and the difference rests with parameters instead of mathematical expression. A mathematical model reflects the structure at the macro level, while the parameter value reflects the rules of evolution or the interaction property of elements at the micro level. Where the macro structure is concerned, there exists significant similarity between the cities in developed countries and those in developing countries. However, where the micro interaction and the evolution rules are concerned, there is clear difference between two types of cities. In fact, it can be demonstrated that the fractal dimension curves of urban form are consistent with the urbanization curve of a country.

The shortcoming of this study lies in empirical analysis. No long time series of fractal dimension of urban form in developed countries is available for the positive study of sigmoid curves of fractal dimension. The measurement and calculation of fractal parameters depend on the definition of cities, the identification of urban boundaries, determination of study area, quality of images or digital maps, selection of algorithms, and so on. In order to testify the logistic function of fractal dimension growth, the caliber of spatial data, definition of study area, and the length of fractal dimension series are very important. However, all these problems cannot be controlled by author in this study. Comparably speaking, we have a good sample path of fractal dimension values of Beijing city to verify the quadratic logistic function of fractal dimension growth of urban form in developing countries (Chen and Huang, 2016).

## 5. Conclusions

The academic contributions of this paper to urban science rest with three aspects. The first is that two spatial measures are proposed for urban description; the second is that a new model of spatial dynamics is constructed by means of the two measures; and third is the strict derivation of the logistic function of fractal dimension curve from the spatial dynamics model. Based on the mathematical deduction, empirical analysis, and numerical experiment, the principal conclusions can be reached as follows. **First, spatial entropy and information gain can serve as a pair of basic spatial measurements of cities.** Urban growth is process of space filling, which can be measured by spatial entropy. For urban form, spatial entropy is a measure of the unavailability of a city's land resource. An increase in spatial entropy suggests a decrease in land availability. If there were no resistance to spatial entropy increase, a city's space would be completely filled and become a kind of Euclidean space. The information gain indicates another force that opposes or even restrains spatial entropy. The interaction between spatial entropy and information gain leads to unity of opposites, and fractal patterns of cities emerge from the paradoxical movement. **Second, the urban-rural interaction model indicative of replacement dynamics can be built by means of spatial entropy and information gain.** Entropy can be associated with fractal dimension by functional box-counting method. In terms of the relation between entropy and fractal dimension, spatial entropy can be used for a measure of urban space filling. Accordingly, the information gain can be used for a measure of urban space saving or even rural space filling. Thus, a system of nonlinear differential equations can be constructed for urban-rural interaction of metropolitan area. The fractal structure of cities implies the spatial order emerging between spatial entropy increase and information generation. **Third, the logistic function of fractal dimension curve can be exactly derived from the urban-rural interaction model based on spatial entropy and information gain.** This mathematical derivation is graceful and revealing. A sigmoid function corresponds to a 1-dimensional map, while the urban-rural interaction models correspond to a 2-dimensional map. The 2-dimensional maps of urban evolution reflect the spatial replacement dynamics. Based on the relationship between the sigmoid functions and the urban-rural interaction models, the 2-dimensional spatial replacement dynamics can be analyzed by the 1-dimensional map based on sigmoid function. Thus, spatial complexity and heterogeneity of cities can be interpreted by the unity of opposites between spatial entropy and information gain. In this sense, the sigmoid functions of fractal dimension curves of urban growth suggest a good approach to exploring spatial dynamics of urban evolution. It is hard to clarify too many questions in a few lines of words, the remaining problems will be solved in future studies.

## Acknowledgements

This research was sponsored by the National Natural Science Foundation of China (Grant No. 41671167). The support is gratefully acknowledged.

# Appendix 1—Modeling fractal dimension curves of London's urban growth using the 2-dimensional map (1786-2020)

Using the 2-dimensional map based on the urban-rural interaction model, we can generate the values of capacity dimension, information dimension, and correlation dimension from 1786 to 2020. The relationships between scatter points indicative of observed values and trend lines indicative of predicted values can be illustrated in Figure A.

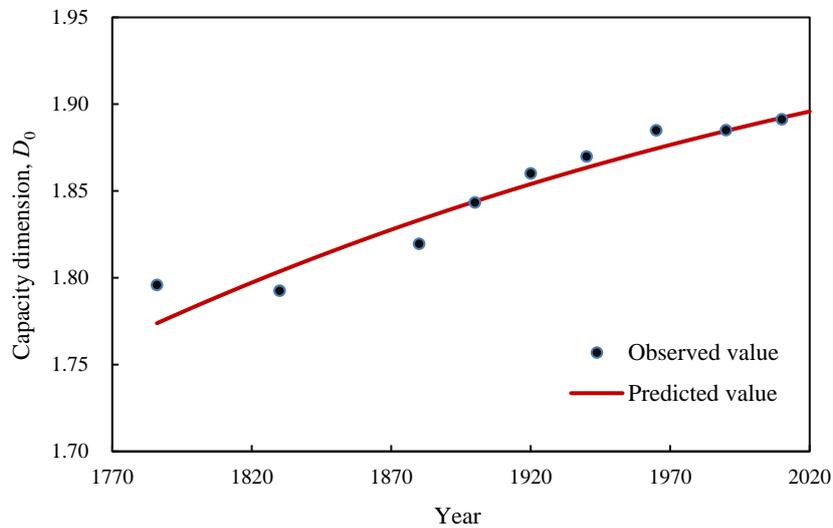

a. Capacity dimension ($1.75<D_0<2.0$)

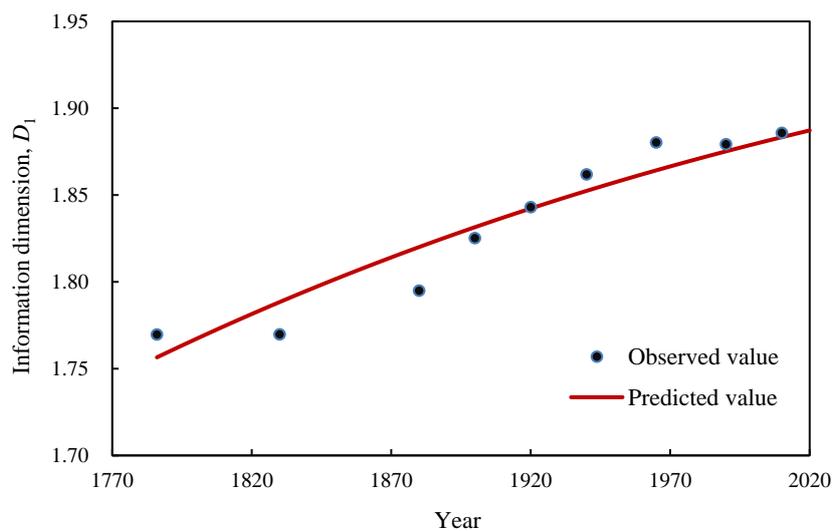

b. Information dimension ($1.7<D_1<2$)

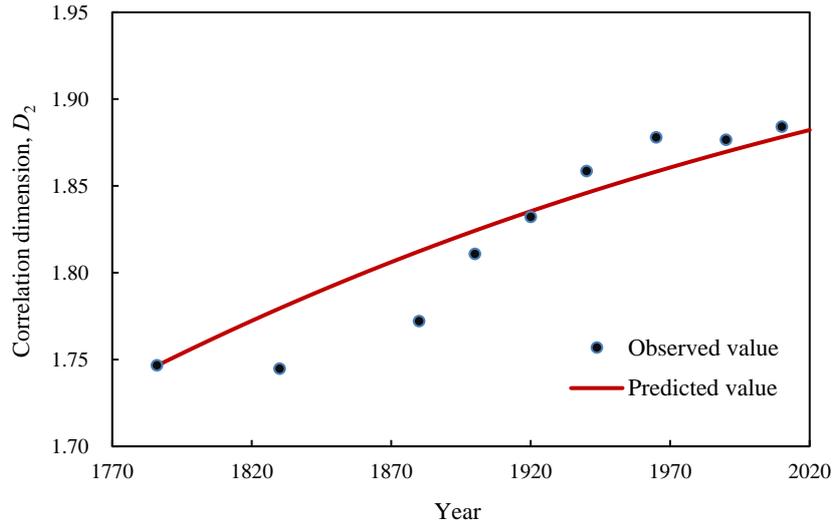

c. Correlation dimension (1.7<$D_2$<2)

**Figure A Modeling three logistic curves of fractal dimension growth of London's urban form with 2-dimensional map (1786-2020)**

# Appendix 2—Quadratic logistic curve of urban fractal dimension growth

A typical city of developing countries is Beijing, the capital of China. The multifractal dimension spectrums have been worked out by Chen and Huang (2016). The quadratic logistic function can be used to model the fractal dimension curves of Beijing's urban form and growth. For capacity dimension ($q=0$), the growth model is

$$\hat{D}_0(t) = \frac{1.9171}{1+0.2778e^{-(0.0626t)^2}}.$$

The goodness of fit is about $R^2=0.9811$. For information dimension ($q=1$), the growth model is

$$\hat{D}_1(t) = \frac{1.8431}{1+0.3088e^{-(0.0639t)^2}}.$$

The goodness of fit is about $R^2=0.9846$. For correlation dimension ($q=2$), the growth model is

$$\hat{D}_2(t) = \frac{1.8097}{1+0.3173e^{-(0.0642t)^2}}.$$

The goodness of fit is about $R^2=0.9870$. Three fractal dimension curves are as below (Figure B).

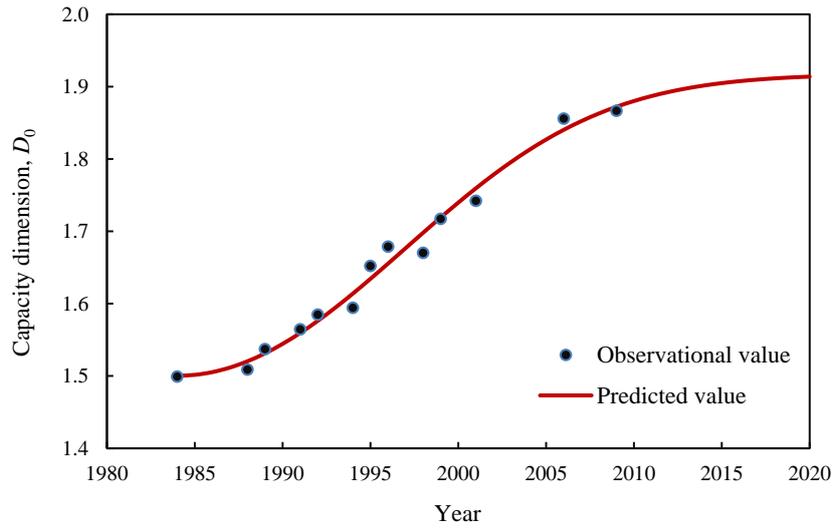
a. Capacity dimension ($1.4<D_0<2.0$)

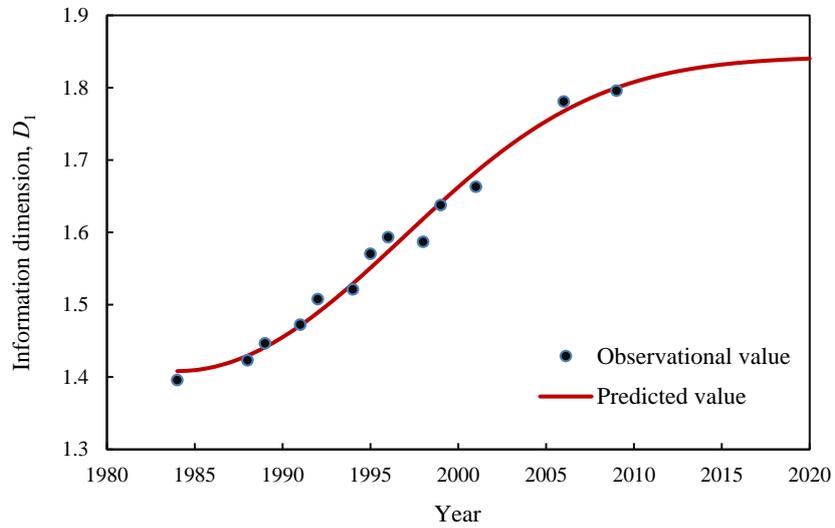
b. Information dimension ($1.3<D_1<1.9$)

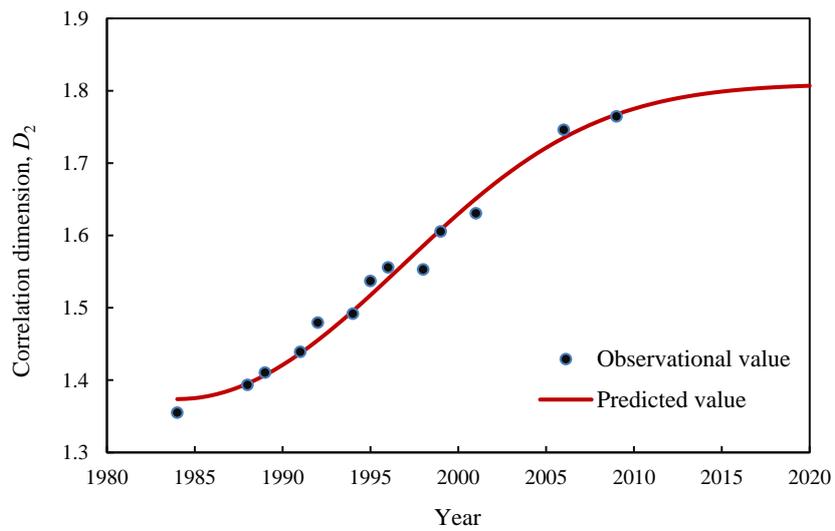
c. Correlation dimension ($1.3<D_2<1.9$)

**Figure B Three quadratic logistic curves of fractal dimension growth of Beijing's urban form (1984-2020) [by Chen and Huang, 2016]**